\documentclass[conference]{IEEEtran}

\ifCLASSINFOpdf

\else
\fi

\usepackage{url}
\usepackage[british]{babel}
\usepackage{hhline}
\usepackage{multirow}
\usepackage{graphicx}
\usepackage{epsfig}
\usepackage{ifmtarg}
\usepackage[font=small,labelsep=none]{caption}
\usepackage[font=footnotesize]{caption}
\usepackage{footnote}
\usepackage{cite}
\usepackage{amssymb}
\usepackage{amsthm}
\usepackage[fleqn]{amsmath}
\usepackage{subfig}
\usepackage{mathtools}
\usepackage{amsfonts}
\usepackage{epstopdf}
\usepackage{epsfig}
\usepackage[linesnumbered,ruled]{algorithm2e}
\usepackage[font=small,labelsep=none]{caption}
\usepackage[font=footnotesize]{caption}
\begin{document}
\title{Hybrid Cell Outage Compensation in 5G Networks: Sky-Ground Approach}
\author{
\IEEEauthorblockN{\large Mohamed Y. Selim, Ahmad Alsharoa and Ahmed E. Kamal} \\
\vspace{-.15 in}
\small
\IEEEauthorblockA{Iowa State University, Iowa State, USA, Email: \{myoussef, alsharoa, kamal\}@iastate.edu}\\ \vspace{-.15 in}

\vspace{-.15 in}
\vspace{-.15 in}
\vspace{-.15 in}
\vspace{-.15 in}
}

\maketitle

\begin{abstract}
%\boldmath
Unmanned Aerial Vehicles (UAVs) enabled communications is a novel and attractive area of research in cellular communications. It provides several degrees of freedom in time, space and it can be used for multiple purposes. This is why wide deployment of UAVs has the potential to be integrated in the upcoming 5G standard. In this paper, we present a novel cell outage compensation (COC) framework to mitigate the effect of the failure of any outdoor Base Station (BS) in 5G networks. Within our framework, the outage compensation is done with the assistance of sky BSs (UAVs) and Ground BSs (GBSs). An optimization problem is formulated to jointly minimize the energy of the Drone BSs (DBSs) and GBSs involved in the healing process which accordingly will minimize the number of DBSs and determine their optimal 2D positions. In addition, the DBSs will mainly heal the users that the GBS cannot heal due to capacity issues. Simulation results show that the proposed hybrid approach outperforms the conventional COC approach. Moreover, all users receive the minimum quality of service in addition to minimizing the UAVs' consumed energy.

\end{abstract}
\begin{IEEEkeywords}
Self-healing, Cell Outage Compensation, Drone-based Communications, Unmanned Aerial Behicles (UAVs), 5G.
\end{IEEEkeywords}
\IEEEpeerreviewmaketitle
\vspace{-.15 in}
\vspace{-.05 in}
\section{Introduction}

Unmanned aerial vehicles (UAVs) enabled communications have attracted considerable attention recently due to the inherent agility \cite{RuiZhangComm}. On demand UAVs can rapidly provide network iaccess to be used in various applications. As reported by AIAA (www.aiaa.org), the global market for commercial UAV applications will skyrocket to as much as 127 billion dollars by 2020. UAVs are gaining increasing popularity in Information Technology (IT) applications due to their high flexibility for on-demand deployments. According to Nokia (www.nokia.com), in May 2016, they launched a rapidly deployable 4G solution that can be carried by drones to provide connectivity at high-traffic events. Also, project Loon by Google (www.google.com/loon) provided internet access worldwide by leveraging the UAV technology. There are other projects led by other companies such as Facebook and AT$\&$T.

In particular, employing UAVs as aerial Base-Stations (BSs) is envisioned as a promising solution to tackle the challenges facing the existing 4G, and forthcoming 5G networks. One of the main challenges facing these networks is the failure of BSs and how to self-heal or mitigate this failure in an autonomous way. A Self Organizing Network (SON) aims to leapfrog the overall performance of the network. SON defines three areas: self-configuration, self-optimization and self-healing \cite{Imran}.

Self-healing is the execution of actions that keep the network operational and/or prevent disruptive problems from arising. Self-healing or specifically Cell Outage Compensation (COC) executes actions to mitigate or, at least, alleviate the effect of the failure \cite{SelimComm}.

When a failure occurs at any BS in the network, the conventional and well-known cell outage compensation technique changes the neighboring BSs' antennae tilt and transmission power levels to serve the users of the failed BS. The advantage of this self-healing technique is that it is fast and guarantees minimum Quality of Service to the users under the failed BS. However, its disadvantage is that the users of the neighboring BSs will be affected by the change in their BSs' antennae configurations.

To make use of the advantage of the conventional self-healing technique and avoid its disadvantage, we propose a novel approach where DBSs will serve users that are not connected to any neighboring Ground BS (GBS) or those users that overloading neighboring GBS and affecting its original users. The proposed approach aims to minimize the number of used DBSs by minimizing the energy of the healing process which consequently minimizes the number of used DBSs and ensures that each user is attached to at least one BS (DBS or GBS) and is receiving the minimum required achievable rate.

Although there has been significant amount of work on using DBSs in cellular networks, using DBSs in self-healing is still at its infancy. In \cite{Rohde}, the positioning of aerial relays is discussed to compensate cell outage and cell overload. The authors in \cite{Merwaday} show the improvement in the coverage by assisting the network with DBSs at a certain altitude, in case of failure of the network BS.

In \cite{Mozaffari}, the optimal altitude of a DBS that
achieves a required coverage with minimum transmission power
is found. Also providing maximum coverage with two DBSs
in the presence and absence of interference is investigated. The authors in \cite{RuiZhangOff} designed an offloading scheme using UAVs where UAVs flies cyclically along the cell edge to serve cell-edge users and help offload data from the GBS.

%The rest of this paper is organized as follows. Section II introduces the system model. In Section III, the optimization problem and its solution is presented. Section IV presents the numerical results. Finally, Section V conclude the paper.

\begin{figure*}[!htb]
    \centering
    \begin{minipage}{3.25in}
        \centerline{\includegraphics[width=3.5in, height=1.7in]{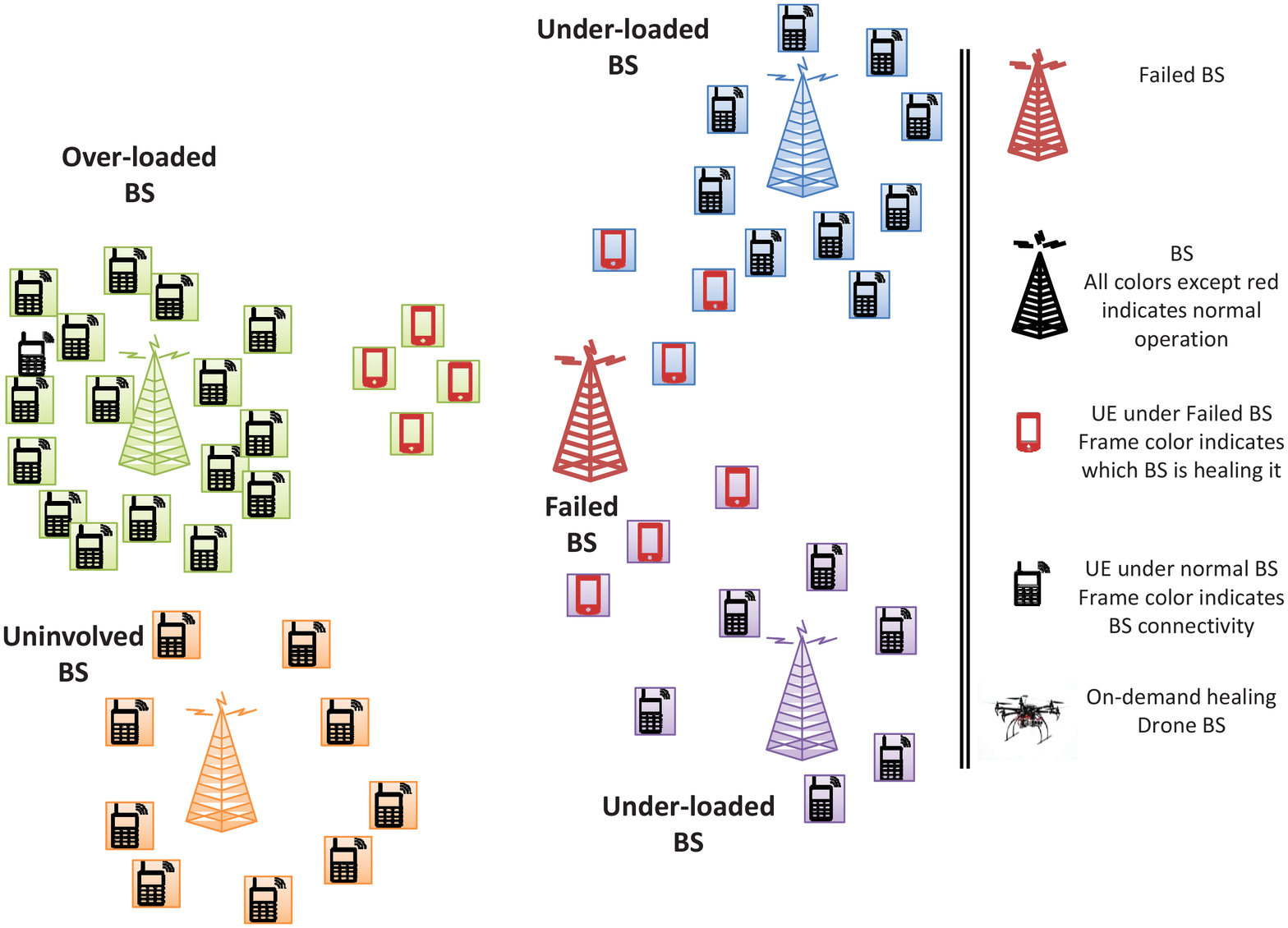}}
i\small
        \caption{\,: Conventional self-healing approach.}
\normalsize
        \label{convs}
    \end{minipage}%
    \hspace*{0.1in}
    \begin{minipage}{3.15in}
        \centerline{\includegraphics[width=3.2in, height=1.7in]{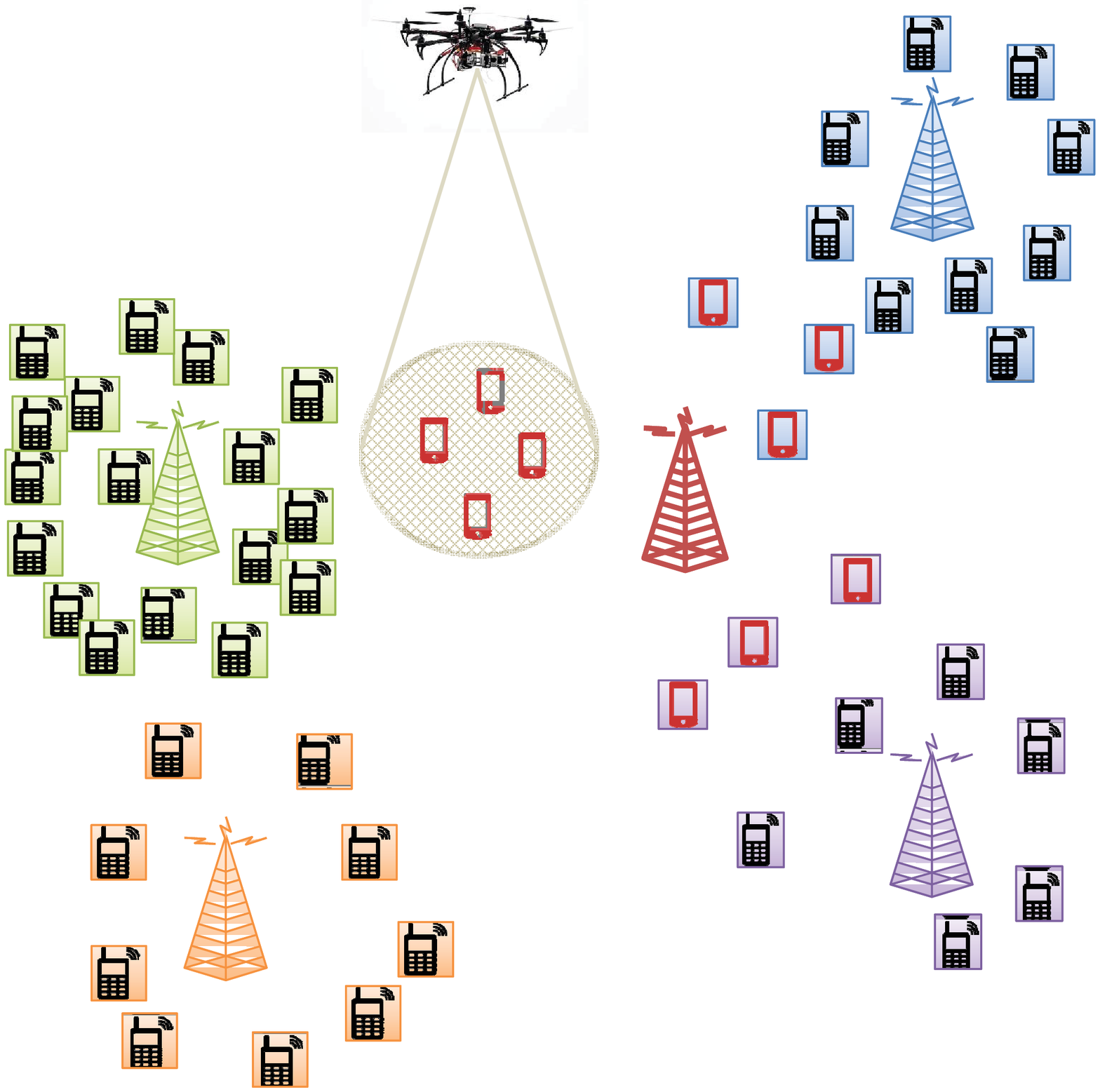}}
\small
        \caption{\,: Hybrid self-healing approach.}
\normalsize
        \label{convswithdrone}
    \end{minipage}%
\vspace{-.15 in}
\end{figure*}

\section{System Model}

As shown in Fig. \ref{convs}, we consider a wireless communication system with a heterogeneous network and $D$ DBSs which are employed to heal a group of $U$ UEs under the failed GBS given one failure at a time or multiple failures in different geographical locations.

The set $\mathcal{U}=\{1,2,\dots,U\}$ denotes the set of active UEs under the failed BS and they are at known locations where the horizontal coordinate of each UE $u$ is fixed at $\mathbf{g}_u=[x_u,y_u]^T$, $u \in U$. The set $\mathcal{D}=\{1,2,\dots,D\}$ denotes the set of DBSs used to heal the failed BS where all DBSs are assumed to navigate at a fixed altitude $h$ and the horizontal coordinate of DBS $d \in D$ at discrete time instant $n$ where $n=1,...,N$ is denoted by $\mathbf{J}_d^n=[x_d^n,y_d^n]^T$ where $N$ is a total discrete period.

We denote that DBS $d$ is used in time block $n$ by $\kappa_d^n$ which acts as a decision variable in our formulation. The UEs under the failed BS are associated with either a DBS or a GBS. We denote $\zeta_{u,d,m}^n$ as the binary variable which indicates that UE $u$ is associated with DBS $d$ and using sub-channel $m$ during time block $n$. Similarly, $\epsilon_{u,l,m}^n$ is defined for GBS $l$.

Assume that the DBS-UE communication channels are dominated by LoS links. Though simplified, the LoS model offers a good approximation for practical DBS-UE channels and enables us to investigate the main objective of the optimization problem presented later. Under the LoS model, the DBS-UE channel power gain follows the free space path loss model. Given that $\mathbf{J}_d^n$, $\mathbf{J}_l$ and $g_u$ as the coordinates of DBS $d$, GBS $j$ and UE $u$ at discrete time instant $n$, respectively, then the distance from DBS $d$ to UE $u$ in time block $n$ can be expressed as:

\small
\begin{equation}\label{dist}
  \delta_{u,d}^n=\sqrt{h_d^2+||\mathbf{J}_d^n-\mathbf{g}_u||^2}
\end{equation}
\normalsize

Similarly, the distance from GBS $l$ to UE $u$ in time block $n$ can be expressed as:

\small
\begin{equation}\label{dist}
  \delta_{u,l}=\sqrt{h_l^2+||\mathbf{J}_l-\mathbf{g}_u||^2}
\end{equation}
\normalsize

where $\mathbf{J}_l$ is constant, $h_l$ is the height of the GBS.

\subsection{DBS and GBS Channel and Achievable Rate Models}

Under this LoS model, the DBS-UE channel power gain is given as follows:

\small
\begin{equation}\label{pathlssd}
  \Gamma_{u,d}^n=\rho_{o} (\delta_0/\delta_{u,d}^n)^{2}=\frac{\rho_{o}}{h^2+||\mathbf{J}_d^n-\mathbf{g}_u||^2}
\end{equation}
\normalsize

where $\rho_{o}$ is a unitless constant, measured at the reference distance $\delta_0=\text{1~m}$, that depends on the antenna characteristics and frequency. Moreover, the channel gain for the communication links between GBS-UE is following the urban path loss model which is given by:

\small
\begin{equation}\label{pathlssl}
  \Gamma_{u,l}=\rho_{o} (\delta_0/\delta_{u,l})^{\alpha}=\frac{\rho_{o}}{\big(\sqrt{h_l^2+||\mathbf{J}_l-\mathbf{g}_u||^2}\big)^\alpha}
\end{equation}
\normalsize

% remember \alpha is 2 for free space and 3 for urban path loss (simulation parameters)

Let $\mathcal{M}=\{1,2,\dots,M\}$ be the set of self-healing sub-channels that each DBS and GBS can use during the self-healing process. These sub-channels will be further divided and allocated to the UEs associated with each DBS and GBS. We assume that each DBS $d$ and GBS $l$ transmits with a constant per sub-channel transmit power $p_{d,m}$ and $p_{l,m}$, respectively. If sub-channel $m$ is not assigned to DBS $d$ then $p_{d,m}$ will be zero. For simplicity, we assume that there is no interference between the DBS tier and the GBS tier. Hence, the received Signal to Interference plus Noise Ratio (SINR) between DBS $d$ and UE $u$ per sub-channel $m$ during time block $n$ can be expressed as:

\small
\begin{equation}\label{SINRd}
  \gamma_{u,d,m}^n=\frac{p_{d,m}~\Gamma_{u,d}^n}{\sum\limits_{\substack{j\in\mathcal{D} \\ j\neq d}} p_{j,m} \Gamma_{u,j}^n + \sigma^2}
  =\frac{\frac{p_{d,m ~\rho_{o}}}{h^2+||\mathbf{J}_d^n-\mathbf{g}_u||^2}}{\sum\limits_{\substack{j\in\mathcal{D} \\ j\neq d}} \frac{p_{j,m} ~\rho_{o}}{h^2+||\mathbf{J}_j^n-\mathbf{g}_u||^2}  + \sigma^2}
\end{equation}
\normalsize

Similarly, we can express the received SINR between GBS $l$ and UE $u$ per sub-channel $m$ during time block $n$ as:

\small
\begin{equation}\label{SINRl}
  \gamma_{u,l,m}=\frac{p_{l,m}~\Gamma_{u,l}}{\sum\limits_{\substack{i\in\mathcal{L} \\ i\neq l}} p_{i,m} \Gamma_{u,i} + \sigma^2}
  =\frac{\frac{p_{l,m}~\rho_{o}}{\big(\sqrt{h_l^2+||\mathbf{J}_l-\mathbf{g}_u||^2}\big)^\alpha}}{\sum\limits_{\substack{i\in\mathcal{L} \\ i\neq l}} \frac{p_{i,m}~\rho_{o}}{\big(\sqrt{h_l^2+||\mathbf{J}_l-\mathbf{g}_u||^2}\big)^\alpha} + \sigma^2}
\end{equation}
\normalsize
%define epsilon and zeta
where $\sigma^2$ is the power of the Additive White Gaussian Noise at the receiver. The first term in the denominator of equations (\ref{SINRd}) and (\ref{SINRl}) represents the co-channel interference caused by the transmissions of all other DBS/GBSs on the same sub-channel $m$, respectively. Thus the achievable per sub-channel rate of UE $u$  connected to DBS $d$ during time block $n$ is $R_{u,d,m}^n= \text{log}_\text{2}(1+\gamma_{u,d,m}^n)$  bps/Hz. Moreover, the achievable per sub-channel rate of UE $u$ connected to GBS $l$ during time block $n$ is $R_{u,l,m}= \text{log}_\text{2}(1+\gamma_{u,l,m})$ bps/Hz.

\subsection{Base Stations Power Model}
In order for any GBS to serve its connected users during a time block $n$, GBS $l$ consumes a certain amount of power. This amount of power can be expressed as ~\cite{EARTH1}:
\small
\begin{equation}\label{BSpowermodel}
  P_l^{n(noSH)}= \alpha_l P_{U_N}+\beta_l,
\end{equation}
\normalsize
where $\alpha_l$ is the scaling parameter, $U_N$ is the total number of users served by the GBS, $P_{U_N}$ is the total power used by this GBS to serve all its users during normal operation and $\beta_l$ models a constant power which is consumed independently of the radiated power of GBS.

Upon the failure of any BS, the neighboring GBSs will heal the users under the failed BS by applying the conventional self-healing approach, i.e., changing antenna tilt and power. Then the additional power consumed by neighboring GBS $l$ during the self-healing period $P_{l}^{n(SH)}$ is the power radiated to heal those users. This can be expressed as:

\begin{equation}\label{BSpowerSH}
  P_{l,m}^{n(SH)}= \tilde{\alpha_l} \sum_{u=1}^{U} \epsilon_{u,l,m}^n p_{l,m}
\end{equation}

where $\tilde{\alpha_l}$ is a scaling parameter which takes into consideration the increase of the BS antenna power during the healing process where $\tilde{\alpha_l} \ge \alpha_l$. $\epsilon_{u,l,m}^n$ is a binary variable indicating the association of the user $u$ with BS $l$ using sub-channel $m$ and $p_{l,m}$ is the fixed amount of power radiated from the GBS to each user connected to it. Note that the additional independent power $\beta_l$ is not accounted in the case of failure since this power is already consumed whether there is a failure or no and in Eq. (\ref{BSpowerSH}) we are only considering the excess consumed power due to the healing process.

\subsection{Drone Power Model}

There are three sources draining power from the DBS battery: 1) The hardware power 2) The hovering power 3) The DBS transmission power. We assume that all drones move with a fixed speed denoted by $v_d$. The hovering and hardware drone power levels, denoted by $P_\text{hov}$ and $P_\text{har}$, can be expressed, respectively, as \cite{Sharoa}:

\small
\begin{equation}
P_\text{hov}= \sqrt{(m_\text{tot} g)^3 / 2 \pi r_p^2 n_p \rho},
\end{equation}
\normalsize

\small
\begin{equation}\label{Ps}
P_\text{har}= \frac{P_\text{full}-P_s}{v_\text{max}}v_d+P_s,
\end{equation}
\normalsize
where $m_\text{tot}$, $g$, and $\rho$ are the drone mass in ($\text{Kg}$), earth gravity in (m$/\text{s}^2$), and air density in $(\text{Kg}/\text{m}^3)$, respectively. $r_p$ and $n_p$ are the radius and the number of the drone's propellers, respectively. $v_\text{max}$ is the maximum speed of the drone and in our model it is equal to $v_d$. $P_\text{full}$ and $P_s$ are the hardware power levels when the drone is moving at full speed and when the drone is in idle mode, respectively. When the DBS is flying to a destination, it will consume $P_\text{har}$. Finally, the total flying power of DBS $d$ can be calculated as $P_f= P_\text{hov}+P_\text{har}$.

The DBS transmission power can be modeled exactly in the same way of the regular BS with the new parameters $\alpha_d$ and $\beta_d$. This can be seen in the second term of Eq. (\ref{DBSenergy}).

\section{Problem Formulation and Proposed Solution}

In this section, we formulate an optimization problem aiming to minimize the total energy of GBSs and DBSs during the hybrid self-healing mechanism which will determine when to use the healing DBSs given capacity and rate constraints. The optimization problem starts after the detection of the failure and hence applying the conventional self-healing technique to serve the affected UEs. Once all UEs are served by the neighboring GBSs, the optimization problem will work mainly on minimizing the overall system energy in addition to serving users of the failed GBS, hence minimizing the number of DBSs used in the healing process.

The total energy consumed by BS $l$ to heal the UEs of the failed BS during time block $n$ is given by the total duration of healing $T$ multiplied by the healing power as follows:
\small
\begin{equation}\label{BSenergy}
  E_{l,m}^n=T P_{l}^{n(SH)}=T \tilde{\alpha_l} \sum_{u=1}^{U} \epsilon_{u,l,m}^n p_{l,m}
\end{equation}
\normalsize

The total energy consumed by any DBS $d$ to heal the users of the failed BS is given by:

\small
\begin{equation}\label{DBSenergy}
E_{d,m}^n= \kappa_d^n\big( T_f(P_\text{har})+ T(P_\text{har}+P_\text{hov})\big)+ T [\alpha_d \sum_{u=1}^{U} \zeta_{u,d,m}^n ~p_{d,m}+\beta_d]
\end{equation}
\normalsize

where $\kappa_d^n$ is a binary variable indicating whether or not DBS $d$ is used in time block $n$ and $T_f$ is the time the DBS takes to travel from its initial position to the position from which it will serve the users.

\subsection{Problem Formulation}
The optimization problem minimizing the energy of the healing BSs (ground and sky BSs) to heal the UEs under the failed OSC is given by:

\small
\begin{align}
&\hspace{-0.5cm}(\bf{P1}):\underset{\mathbf{J}_d^n, {\epsilon_{\textit{u,l,m}}^\textit{n}}, {\zeta_{\textit{u,d,m}}^\textit{n}}, {\kappa_\textit{d}^\textit{n}}}{\text{minimize}} \sum_{\textit{n=1}}^{\textit{N}}\sum_{\textit{l=1}}^{\textit{L}} \sum_{\textit{m=1}}^{\textit{M}} \textit{E}_\textit{l,m}^\textit{n}+ \sum_{\textit{n=1}}^{\textit{N}}\sum_{\textit{d=1}}^{\textit{D}} \sum_{\textit{m=1}}^{\textit{M}} \textit{E}_\textit{d,m}^\textit{n}    \label{of}\\
&\hspace{-0.5cm}\text{subject to:}\nonumber\\
&\hspace{-0.5cm} \sum_{l=1}^{L}\sum_{m=1}^{M}\epsilon_{u,l,m}^n + \sum_{d=1}^{D}\sum_{m=1}^{M} \zeta_{u,d,m}^n = 1   ~~~\forall~ u,n                    \label{AssociationC}\\
&\hspace{-0.5cm} \sum_{d=1}^{D}\sum_{m=1}^{M} \zeta_{u,d,m}^n R^n_{u,d,m} + \sum_{l=1}^{L}\sum_{m=1}^{M} \epsilon_{u,l,m}^n R_{u,l,m} \geq R^{\text{th}}_u~~~\forall~ u,n \label{RateConstraint}\\
&\hspace{-0.5cm} R^{th}_u(\sum_{u=1}^{U} \sum_{m=1}^{M} \epsilon_{u,l,m}^n + \overline{U}_l^n) \leq R^{\text{max}}_{GBS}   ~~~\forall~ l,n            \label{Ratebackhaul}\\
&\hspace{-0.5cm} R^{th}_u(\sum_{u=1}^{U}\sum_{m=1}^{M} \zeta_{u,d,m}^n) \leq R^{\text{max}}_{DBS}   ~~~\forall~ d,n                                  \label{backhaul}\\
&\hspace{-0.5cm} \kappa_d^n < 1+\frac{\sum_{u=1}^{U}\sum_{m=1}^{M}\zeta_{u,d,m}^n}{Q}   ~~~\forall ~d,n                                              \label{kappa1}\\
&\hspace{-0.5cm} \kappa_d^n \geq \frac{\sum_{u=1}^{U}\sum_{m=1}^{M} \zeta_{u,d,m}^n}{Q} ~~~\forall ~d,n                                              \label{kappa2}\\%Q very large number
&\hspace{-0.5cm}\mathbf{J}_d^{\text{min}} \leq \mathbf{J}_d^n \leq \mathbf{J}_d^{\text{max}}, \quad \forall~ d,n                                 \label{Coordinates}\\
&\hspace{-0.5cm} \kappa_d^n, \zeta_{u,d,m}^n, \epsilon_{u,l,m}^n \in \{0,1\}                                                                     \label{binaryC}
\end{align}
\normalsize

Constraint (\ref{AssociationC}) forces UE $u$ to be associated with DBS $d$ or GBS $l$. Constraint (\ref{RateConstraint}) indicates that the rate of UE $u$, which is associated with either DBS $d$ or GBS $l$, is lower bounded by a threshold rate $R^{th}_u$. Constraints (\ref{Ratebackhaul}) and (\ref{backhaul}) define an upper bound for the maximum rate for GBS and DBS, respectively, given that $\overline{U}_l^n$ is the number of UEs served by GBS $l$ at time block $n$. Since $\kappa_d^n$ indicates whether DBS $d$ is used in time block $n$ or not, constraints (\ref{kappa1}) and (\ref{kappa2}) are used to extract this information from $\zeta_{u,d,m}^n$ where when $\zeta_{u,d,m}^n=\text{0}$ then consequently $\kappa_d^n=\text{0}$ and when $\zeta_{u,d,m}^n=\text{1}$ for any UE $u$ and resource block $M$ then $\kappa_d^n=\text{1}$. Note that $Q$ is a very large number. Constraint (\ref{Coordinates}) is used to limit the 2D coordinates of DBS $d$ where $\mathbf{J}_d^{\text{min}}=[x_d^{\text{min}},y_d^{\text{min}}]^T$ and $\mathbf{J}_d^{\text{max}}=[x_d^{\text{max}},y_d^{\text{max}}]^T$.

$\bf{P1}$ is not easy to solve due to the following: 1) the decision variables $\kappa_d^n$, $\zeta_{u,d,m}^n$, $\epsilon_{u,l,m}^n$ are binary and thus the objective function (\ref{of}) and constraints (\ref{AssociationC})-(\ref{kappa2}) involve integer constraints. 2) Even if we fixed the decision variables, constraint (\ref{RateConstraint}) is still non-convex with respect to DBS coordinates variable $\mathbf{J}_d^n$. Therefore, problem (\ref{of}) is mixed-integer non-linear non-convex problem, which is difficult to be solved optimally.

\subsection{Proposed Solution}

In general, $\bf{P1}$ has no standard method for solving it efficiently. In the following, we propose an efficient iterative algorithm for solving $\bf{P1}$. Specifically, for a given coordinate  $\mathbf{J}_d^n$, we optimize the decision variables, i.e. $\zeta$, $\kappa$ and $\epsilon$, by solving a Linear Program (LP) after relaxing the decision variables. For any given $\zeta$, $\kappa$ and $\epsilon$, the DBS coordinates $\mathbf{J}_d^n$ are optimized based on the Successive Convex Approximation (SCA) technique \cite{SCAICC}. Finally, an iterative algorithm is given to solve $\bf{P1}$ efficiently.

\subsubsection{Solving for Decision Variables}

By fixing the DBS coordinates, the resulting problem will be an Integer LP (ILP) which can be solved optimally but not efficiently due to the large number of binary variables. In this case, relaxing the binary variables and then reconstructing them will allow us to solve this problem efficiently (reconstruction phase is skipped due to space limitation). Hence, for any given $\mathbf{J}_d^n$, the variables of $\bf{P1}$ can be optimized by solving the following problem:

\small
\begin{align}
&\hspace{-0.5cm}(\bf{P2}):\underset{{\epsilon_{\textit{u,l,m}}^\textit{n}}, {\zeta_{\textit{u,d,m}}^\textit{n}}, {\kappa_\textit{d}^\textit{n}}}{\text{minimize}} \quad  \quad \sum_{\textit{n=1}}^{\textit{N}}\sum_{\textit{l=1}}^{\textit{L}} \textit{E}_\textit{l}^\textit{n}+ \sum_{\textit{n=1}}^{\textit{N}}\sum_{\textit{d=1}}^{\textit{D}} \textit{E}_\textit{d}^\textit{n}    \label{of}\\
&\hspace{-0.5cm}\text{subject to:}\nonumber\\
&\hspace{-0.5cm} \quad \quad \quad \quad \quad \text{Constraints} (\ref{AssociationC})-(\ref{kappa2})                                   \nonumber\\
&\hspace{-0.5cm} 0 \leq \kappa_d^n, \zeta_{u,d,m}^n, \epsilon_{u,l,m}^n \leq 1 ~~~~\forall~ u,d,l,n
\end{align}
\normalsize

Note that in $\bf{P2}$, $R_{u,d,m}^n$ is not a variable anymore since we fixed the DBS coordinates. The relaxed $\bf{P2}$ is an LP which can be solved using any LP solver.

\subsubsection{Solving for DBS Coordinates}

For any given decision variable, the DBS coordinates $\mathbf{J}_d^n$ can be optimized by solving the following problem:

\small
\begin{align}
&\hspace{-0.5cm}(\bf{P3}):\underset{\mathbf{J}_d^n}{\text{minimize}} \quad  \quad \sum_{\textit{n=1}}^{\textit{N}}\sum_{\textit{l=1}}^{\textit{L}} \textit{E}_\textit{l}^\textit{n}+ \sum_{\textit{n=1}}^{\textit{N}}\sum_{\textit{d=1}}^{\textit{D}} \textit{E}_\textit{d}^\textit{n}    \label{of}\\
&\hspace{-0.5cm}\text{subject to:}\nonumber\\
&\hspace{-0.5cm} \quad \quad \quad \quad \quad \text{Constraints} (\ref{RateConstraint}), (\ref{Coordinates})                                   \nonumber
\end{align}
\normalsize

In $\bf{P3}$, Constraints (\ref{AssociationC}), (\ref{Ratebackhaul})-(\ref{kappa2}) and (\ref{binaryC}) are not involved in $\bf{P3}$ since the decision variables are now fixed and their values are iteratively updated from $\bf{P2}$. The objective function and all constraints of $\bf{P3}$ are convex except for constraint (\ref{RateConstraint}). This constraint is neither concave nor convex with respect to the DBSs' coordinates which appears in $R_{u,d,m}^n$. It is worth noticing that the second term of the same constraint is not a function of the DBSs' coordinates, hence it is linear. Returning back to the first term of constraint (\ref{RateConstraint}), call it $\tilde{R}$, which can be expanded as follows:

\small
\begin{align}\label{expandR}
  \tilde{R}= & \sum_{d\in\mathcal{D}} \sum_{m\in\mathcal{D}} \zeta_{u,d,m}^n \text{log}_\text{2}(1+\frac{\frac{p_{d,m ~\rho_{o}}}{h^2+||\mathbf{J}_d^n-\mathbf{g}_u||^2}}{\sum\limits_{\substack{j\in\mathcal{D} \\ j\neq d}} \frac{p_{j,m} ~\rho_{o}}{h^2+||\mathbf{J}_j^n-\mathbf{g}_u||^2}  + \sigma^2})  \nonumber\\
    & \sum_{d\in\mathcal{D}} \sum_{m\in\mathcal{D}} \zeta_{u,d,m}^n \text{log}_\text{2}(\frac{\sum_{j\in\mathcal{D}} \frac{p_{j,m}\rho_{o}}{h^2+||\mathbf{J}_j^n-\mathbf{g}_u||^2}+\sigma^2}{\sum\limits_{\substack{j\in\mathcal{D} \\ j\neq d}} \frac{p_{j,m} ~\rho_{o}}{h^2+||\mathbf{J}_j^n-\mathbf{g}_u||^2}  + \sigma^2}) \nonumber\\
  = & \sum_{d\in\mathcal{D}} \sum_{m\in\mathcal{D}} \zeta_{u,d,m}^n \Big( \underbrace{\text{log}_\text{2}(\sum_{j\in\mathcal{D}} \frac{p_{j,m}\rho_{o}}{h^2+||\mathbf{J}_j^n-\mathbf{g}_u||^2}+\sigma^2)}_{\tilde{R}^1}  \nonumber\\
    & \underbrace{-\text{log}_\text{2}(\sum\limits_{\substack{j\in\mathcal{D} \\ j\neq d}} \frac{p_{j,m} ~\rho_{o}}{h^2+||\mathbf{J}_j^n-\mathbf{g}_u||^2}  + \sigma^2)}_{\tilde{R}^2} \Big)
\end{align}
\normalsize

Our main goal is to convert Eq. (\ref{expandR}) to a concave form in order for $\bf{P3}$ to be convex. Both terms of $\tilde{R}$ are neither concave nor convex. $\tilde{R}^2$ is not concave with respect to $\mathbf{J}_j^n$, however, it is concave with respect to $||\mathbf{J}_j^n-\mathbf{g}_u||^2$. This motivates us to introduce the slack variable $\mathbf{\Psi}=\left\{\Psi_{u,j}^n = ||\mathbf{J}_j^n-\mathbf{g}_u||^2 , \forall j \in \mathcal{D}, j\neq d, u,n \right\}$ to make $\tilde{R}^2$ concave in $\mathbf{\Psi}$. After introducing this slack variable to $\tilde{R}^2$, we have to add a new constraint to $\bf{P3}$ which is expressed as \cite{RuiZhangTraj}:

\small
\begin{equation}\label{tildeR^2}
  \Psi_{u,j}^n \leq  ||\mathbf{J}_j^n-\mathbf{g}_u||^2 ~~~~~~~~~~\forall j \in \mathcal{D}, j \neq d, u,n
\end{equation}
\normalsize

Back to the first term of $\tilde{R}$, i.e., $\tilde{R}^1$, this term is neither concave nor convex. Even with the slack variable, $\tilde{R}^1$ will not be concave (it will be convex). To tackle the non-concavity of $\tilde{R}^1$, the SCA technique is applied where in each iteration, the original function is approximated by a more tractable function at a given local point. Define $\mathbf{J}_d^n(r)$ as the given location of DBS $d$ in the r-th iteration. Recall that $\tilde{R}^1$ is convex in $||\mathbf{J}_j^n-\mathbf{g}_u||^2$ and since any convex function can be globally lower-bounded by its first order Taylor expansion \cite{Boyd}, hence, given $\mathbf{J}_d^n(r)$ in iteration r, we obtain the following lower bound:%for $\tilde{R}^1$:

\small
\vspace{-0.5cm}
\begin{align}\label{taylor1st}
  \tilde{R}^1 \geq & ~\text{log}_\text{2}\Big(\sum_{j \in \mathcal{D}}\frac{p_{j,m}\rho_{o}}{h^2+||\mathbf{J}_j^n(r)-\mathbf{g}_u||^2} \Big) \nonumber\\
  - & \sum_{j \in \mathcal{D}} Z_{u,d}^n(||\mathbf{J}_j^n-\mathbf{g}_u||^2 - ||\mathbf{J}_j^n(r)-\mathbf{g}_u||^2) =\tilde{\tilde{R}}^1
%  = & \tilde{\tilde{R}}^1
\end{align}
\normalsize

\vspace{-1.0cm}
\begin{equation}\label{ZzZ}
\hspace{-0.5cm}\text{where} ~ ~ ~ Z_{u,d}^n=\frac{\frac{p_{j,m}\rho_{o}}{h^2+(||\mathbf{J}_j^n(r)-\mathbf{g}_u||^2)^2}\text{log}_\text{2}(e)}{\sum_{k\in\mathcal{D}}\frac{p_{k,m}\rho_{o}}{h^2+||\mathbf{J}_k^n(r)-\mathbf{g}_u||^2}+\sigma^2}
\end{equation}

After using SCA with $\tilde{R}^1$ and using a slack variable with $\tilde{R}^2$, now Eq.(\ref{expandR}) is concave. Hence, with any given local point $\mathbf{J}_j^n(r)$, problem $\bf{P3}$ can be approximated to $\bf{\overline{P3}}$ as follows:

\small
\vspace{-1.0cm}
\begin{align}
&\hspace{-0.5cm}(\bf{\overline{P3}}):\underset{\mathbf{J}_d^n, \Psi_{u,j}^n}{\text{minimize}} \quad  \quad \sum_{\textit{n=1}}^{\textit{N}}\sum_{\textit{l=1}}^{\textit{L}} \textit{E}_\textit{l}^\textit{n}+ \sum_{\textit{n=1}}^{\textit{N}}\sum_{\textit{d=1}}^{\textit{D}} \textit{E}_\textit{d}^\textit{n}    \label{of}\\
&\hspace{-0.5cm}\text{subject to:}\nonumber\\
&\hspace{-0.5cm} \sum_{d=1}^{D} \sum_{m=1}^{M} \zeta_{u,d,m}^n \Big(\tilde{\tilde{R}}^1-\text{log}_\text{2}(\sum\limits_{\substack{j\in\mathcal{D} \\ j\neq d}} \frac{p_{j,m} ~\rho_{o}}{h^2+||\mathbf{J}_j^n-\mathbf{g}_u||^2}  + \sigma^2)\Big)  \nonumber\\
&\hspace{-0.5cm} + \sum_{l=1}^{L}\sum_{m}^{M} \epsilon_{u,l,m}^n R_{u,l,m} \geq R^{\text{th}}_u~~~\forall~ u,n  \\
&\hspace{-0.5cm} \Psi_{u,j}^n \leq  ||\mathbf{J}_j^n-\mathbf{g}_u||^2 ~~~~~~~~~~\forall j \in \mathcal{D}, j \neq d, u,n \\
&\hspace{-0.5cm} \mathbf{J}_d^{\text{min}} \leq \mathbf{J}_d^n \leq \mathbf{J}_d^{\text{max}}, \quad \forall~ d,n
\end{align}
\normalsize

Note that the value of $R_u^\text{th}$ is selected sufficiently low in order to make constraint (30) feasible while taking into consideration inequality (27).

Finally, we propose an iterative algorithm to solve $\bf{P1}$. The variables in $\bf{P1}$ are partitioned into two blocks, i.e., association and coordinates. Then they are alternately optimized (solving $\bf{P2}$ then $\bf{\overline{P3}}$ iteratively). Furthermore, the obtained solution in each iteration is used as the input to the next iteration. The details of this algorithm are summarized in Algorithm 1.

\begin{algorithm}
\small
\caption{Iterative approximate solution for $\bf{P1}$}
\KwIn{$\mathbf{J}_d^n(0)$}
\KwOut{$\mathbf{J}_d^n(r+1),\kappa_d^n(r+1),\zeta_{u,d,m}^n(r+1),\epsilon_{u,l,m}^n(r+1)$}
\While{r $\neq$ maximum iteration}{

Solve Problem $\bf{P2}$ for given $\mathbf{J}_d^n(r)$.

Reconstruct the binary variables, check their feasibility and then denote them as $\kappa_d^n(r+1), ~\zeta_{u,d,m}^n(r+1) ~\text{and}~ \epsilon_{u,l,m}^n(r+1)$

Solve Problem $\bf{\overline{P3}}$ for given $\kappa_d^n(r+1),\zeta_{u,d,m}^n(r+1),\epsilon_{u,l,m}^n(r+1)$.

Denote the optimal solution of $\bf{\overline{P3}}$ as $\mathbf{J}_d^n(r+1)$

Update r=r+1

\If{The fractional increase of the objective value $\leq \varepsilon^{th}$}{Break}
}

\normalsize
\end{algorithm}

\vspace{-0.25in}

\section{Simulation Results}

In this section, numerical results are provided to investigate the benefits of utilizing DBSs in mitigating GBS failures in 5G networks. The simulation model consists of 5 GBSs where  one of them fails. We initialized 4 standby DBSs to be used in case the conventional self-healing approach cannot accommodate the users originally served by the failed BS.

\small
\begin{table}
\vspace{-0.5cm}
\centering
\caption{\label{tab2} System parameters}
\addtolength{\tabcolsep}{-2pt}\begin{tabular}{|l|c||l|c||l|c|}
\hline
\textbf{Parameter}      & \textbf{Value}  & \textbf{Parameter}             & \textbf{Value}  & \textbf{Parameter}      & \textbf{Value}\\ \hline \hline
$f$ (GHz)               &    2.1          & $R^{\text{max}}_{GBS}$ (bps/Hz)&  100            &$\mathbf{J}_d^{\text{min}}$  (m)  & \bf{\text{-200}}   \\ \hline
$p_{d,m}$ (mW)          &    100          & $R^{\text{max}}_{DBS}$ (bps/Hz)&  10             & $\mathbf{J}_d^{\text{max}}$ (m)  & \bf{\text{200}} \\ \hline
$P_{l,m}$ (mW)          &    100          & $R_u^{th}$ (bps/Hz)            &  2              & $h_l$ (W)               & 30 \\ \hline
$\sigma^2$ (dBm/Hz)     &    -174          & $\beta_d$ (W)                  &  1              & $h_d$  (min)            & 100 \\ \hline
$T$ (min)               &    60           & $\alpha_d$                     &  2.6            & $\rho_o$                & 0.01 \\ \hline
$T_f$ (min)             &    0.5          & $\alpha_l$                     &  4.7            & Q                       & 1000 \\ \hline
\end{tabular}
%\vspace{-1.0cm}
\end{table}
\normalsize

\small
\begin{table}
  \centering
  \caption{\label{tab3} Association and rates for 10 UEs}
  \renewcommand{\arraystretch}{1.2}
  \begin{tabular}{|p{1cm}|c|c|c|c|}
    \hline
    \multirow{2}{1cm}{\textbf{UEs}} & \multicolumn{2}{c|}{\textbf{Time block 1 (n=1)}} & \multicolumn{2}{c|}{\textbf{Time block 2 (n=2)}}\\
    % \hline
    % \textbf{Inactive Modes} & \textbf{Description}\\
    \cline{2-5}
    & \textbf{Association} & \textbf{R (bps/Hz)} & \textbf{Association} & \textbf{R (bps/Hz)} \\
    %\hhline{~--}
    \hline
    UE1 & GBS1 & 2.82 & GBS1  &2.28\\ \hline
    UE2 & GBS2 & 2.43 & GBS2  &2.66 \\ \hline
    UE3 & DBS1 & 3.28 & DBS4  &3.59 \\ \hline
    UE4 & GBS3 & 2.41 & GBS3  &2.33\\ \hline
    UE5 & DBS1 & 3.00 & DBS4  &3.37\\ \hline
    UE6 & GBS2 & 2.28 & GBS2  &2.25 \\ \hline
    UE7 & GBS4 & 2.17 & DBS4  &3.12 \\ \hline
    UE8 & GBS3 & 3.07 & GBS3  &2.71\\ \hline
    UE9 & DBS1 & 2.80 & GBS1  &2.30 \\ \hline
    UE10& GBS2 & 2.00 & GBS2  &2.00\\ \hline
  \end{tabular}
\vspace{-0.5cm}
\end{table}
\normalsize

The simulation area is 400$\times$400 $\text{m}^\text{2}$ where the failed GBS is centered at the origin and the UEs of the failed BS are distributed randomly over this area. The UEs of the failed GBS are static, however, the number of users within each neighboring GBS changes randomly per time block. The parameters used to calculate $P_\text{hov}$ and $P_\text{har}$ are initialized as in \cite{Sharoa}. The remaining parameters are presented in Table I.

Table 2 shows the users association (DBS or GBS) and rates for 10 UEs during 2 time blocks. The remaining time blocks are not shown due to space limitations. For time block 1, only DBS 1 is used from a set of 4 DBSs and all other UEs are served by GBSs. Since DBS 1 is serving UE 3, UE5 and UE 9, their corresponding rates are relatively high. UE 1 is associated with the same GBS during time blocks 1 and 2. However, its rate decreases during time block 2. This is due to the change of the capacity of GBS 1 since GBS 1 has to serve its own UEs first and participate in the healing process by the available capacity. During time block 2, DBS 4 is serving UE 3, UE 5 and UE 9. According to these UEs' locations, DBS 4 optimizes its location to serve each of them.

\begin{figure}[!htb]
            \centering
        \includegraphics[width=3in, height=1.825in]{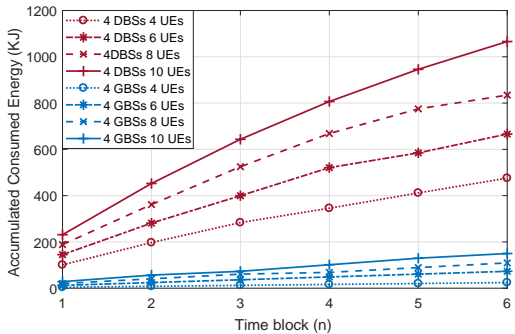}
        \caption{\,: Accumulated energy for DBSs and GBSs.}
        \label{Acc}
\end{figure}

Figure \ref{Acc} shows the accumulated consumed energy for both DBSs and GBSs for different numbers of UEs. It is worth noting that the GBS energy is the excess energy consumed to serve the users originally served by the failed BS. On the other hand, the energy consumed by the DBS is the hovering, hardware and communication power which is significantly high compared with the excess energy consumed by the GBS. As the number of UEs increases from 4 to 10, the number of used DBS is increasing since the GBSs are serving their own UEs and serving the targeted UEs using only the available capacity. At a certain point, all DBSs are used to satisfy the target UEs minimum rate requirement.

\begin{figure*}[!htb]
    \centering
    \begin{minipage}{1.62in}
        \centerline{\includegraphics[width=1.62in, height=1.5in]{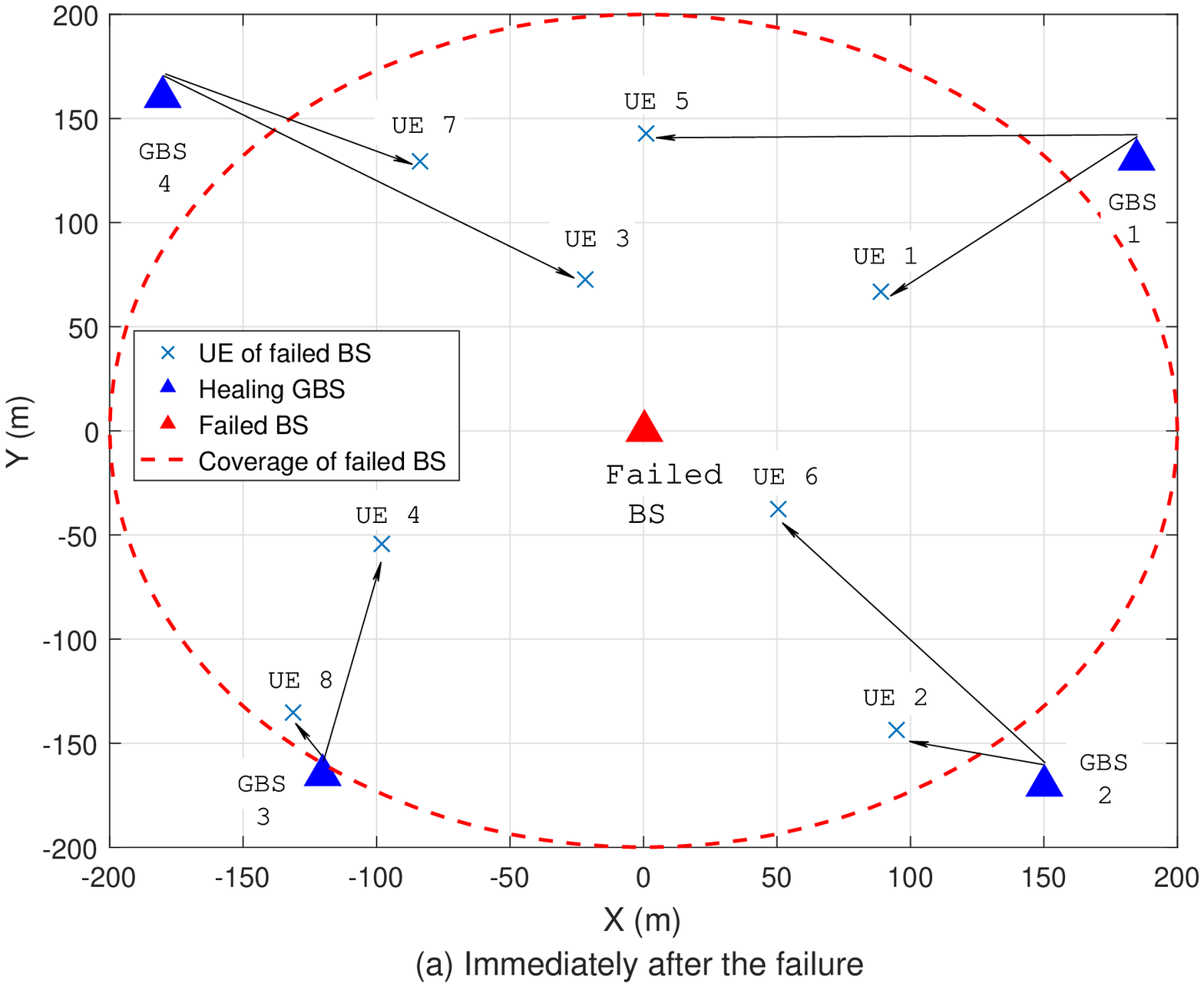}}
    \end{minipage}
    \begin{minipage}{1.62in}
       \centerline{\includegraphics[width=1.62in, height=1.5in]{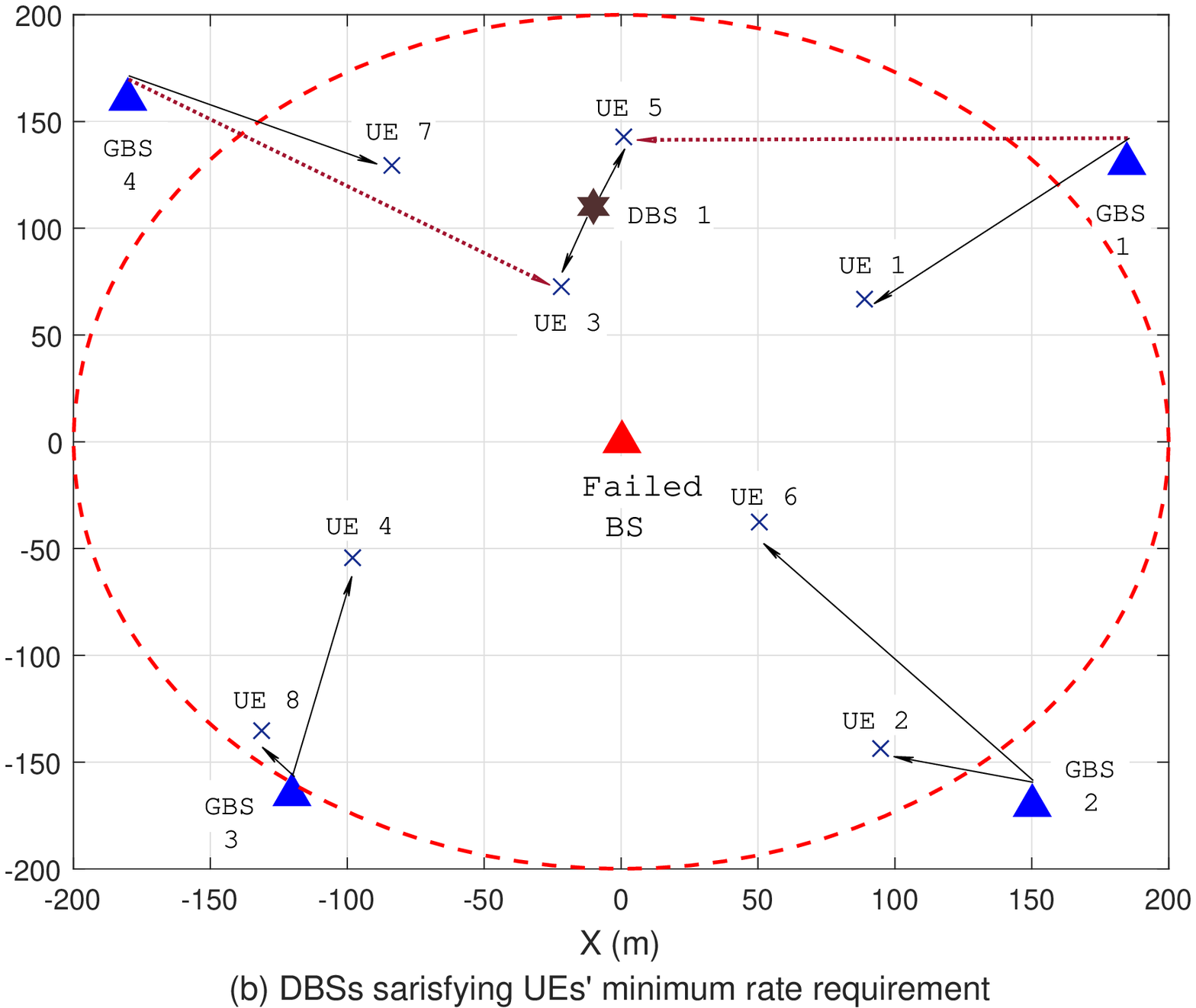}}
    \end{minipage}
    \begin{minipage}{1.62in}
        \centerline{\includegraphics[width=1.62in, height=1.5in]{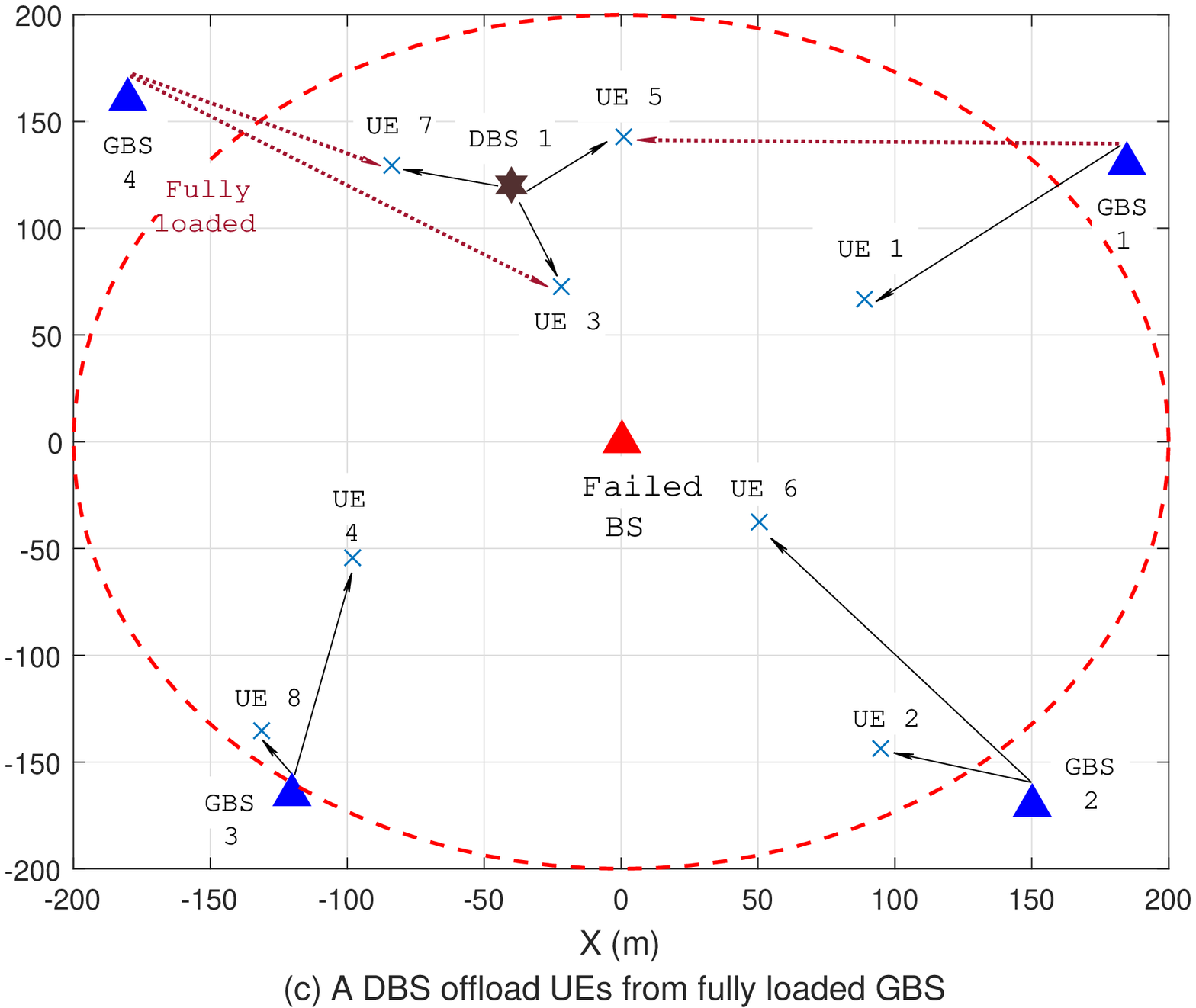}}
    \end{minipage}
    \begin{minipage}{1.62in}
        \centerline{\includegraphics[width=1.62in, height=1.5in]{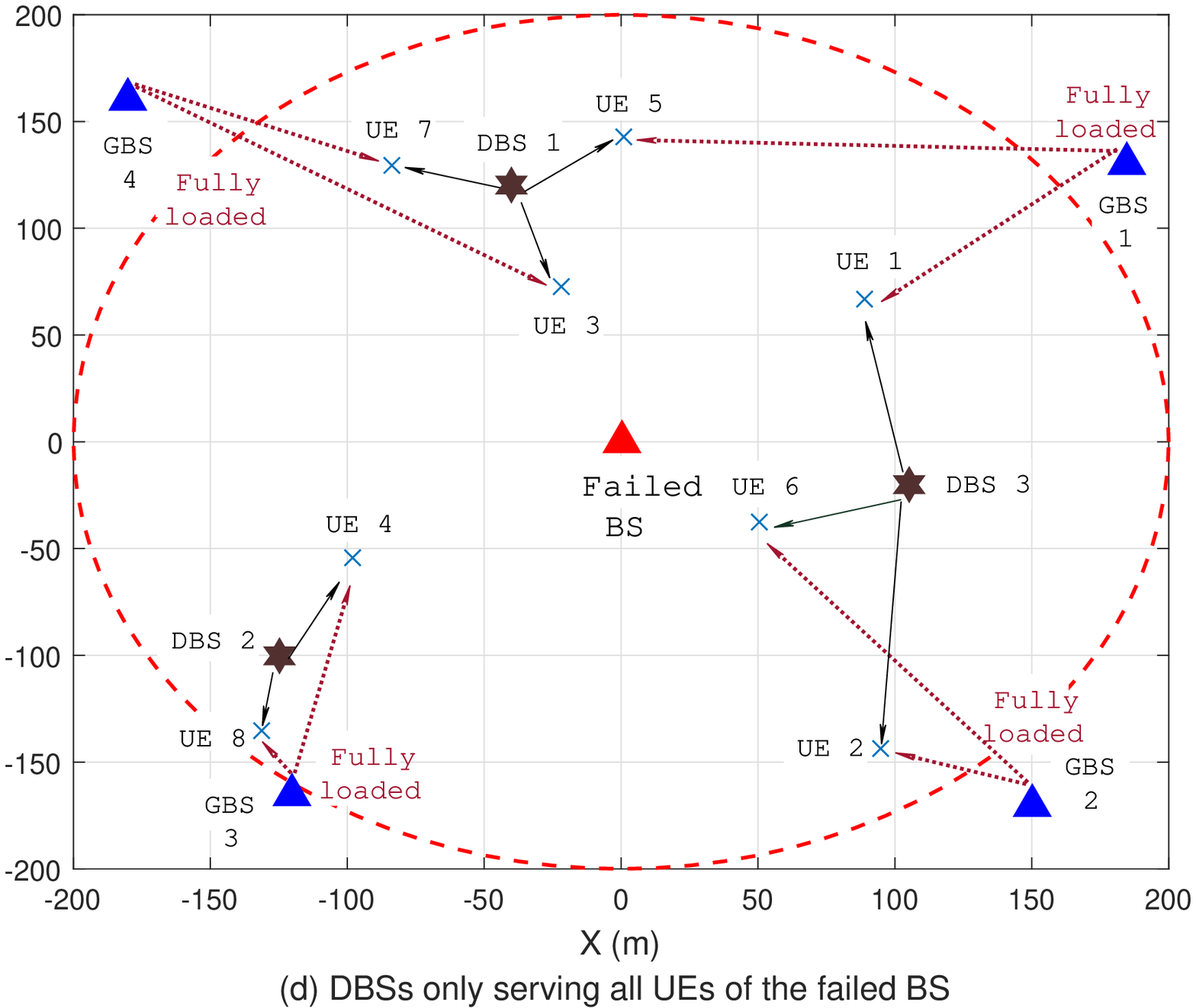}}
    \end{minipage}
\small
    \caption{\,: Cell outage compensation different scenarios (GBSs COC, hybrid COC and DBSs COC) with 4 DBSs, 4 GBSs and 8 UEs.}
\normalsize
\label{DBSGBSx}
\end{figure*}

Fig. \ref{DBSGBSx} shows different scenarios of the proposed scheme where there are 8 UEs connected to the failed BS and there are 4 neighboring GBSs and 4 DBSs ready to participate. In Fig. \ref{DBSGBSx}(a), the GBSs are serving all the UEs without any help from the DBSs. This occurs at the detection of the failure or if the GBSs are non loaded with their own users and they can satisfy all UEs rate requirements. In Fig. \ref{DBSGBSx}(b), UE 3 and UE 5 are not achieving their minimum rate $R_u^{th}$ by associating to GBS 4 and GBS 1, respectively. In this case these two UEs are associated with DBS 1. Although attaching them to DBS 1 will not reduce the energy, this will satisfy UE 3 and UE 7 threshold rates subject to constraint (\ref{RateConstraint}).

Figure \ref{DBSGBSx}(c) shows the scenario of the proposed hybrid algorithm. In this scenario, GBS 4 is fully loaded with its own UEs, hence, UE 7 will be associated with DBS 1 which is already serving UE 3 and UE 5. If DBS 1 was overloaded, then an additional DBS will be used. Finally, Fig. \ref{DBSGBSx}(d) shows a scenario where all GBSs are fully loaded. This scenario is subject to feasibility based on the number of available DBSs. It is worth noting that the 4th DBS was not involved in the healing process in all scenarios.

From the simulation results, we can infer that the hybrid COC is converted to the conventional COC approach if the GBSs are having enough extra capacity. If the number of UEs increases, the disadvantage of the conventional approach will start to appear where either the GBS will not serve the target UE or will degrade the rate of its own UE. Using hybrid approach, we can avoid this scenario by using a DBS to serve those UEs. This confirms that the proposed approach circumvents the disadvantages of the conventional COC approach where the UEs of the neighboring GBSs are not affected by the failure and at the same time the UEs of the failed BS obtain continuous service.

One of the challenges facing this approach is the movement of a DBS from one location to another which is assumed to happen instantaneously in this paper. This challenge can be addressed by adding a velocity constraint to limit the movement of the DBS to its maximum speed.

\section{Conclusion}
In this paper, we proposed a novel cell outage compensation (COC) approach for 5G networks assisted by Drone Base-Stations (DBSs). The objective is to minimize the total energy consumption of the DBSs and Ground BSs (GBSs) while maintaining the minimum quality of service requirements of users originally served by the failed BS. DBSs are optimally managed in order to serve the users that can not be served by GBSs while considering DBSs consumed energy. The simulation results show how this hybrid COC approach outperforms the conventional COC approach. The proposed hybrid approach shows significant impacts on ensuring connectivity of the users originally served by the failed BS while minimizing the number of used DBSs.

\end{document}